\documentclass[aps,prb,twocolumn,showpacs,superscriptaddress,floatfix]{revtex4-1}

\usepackage{amsfonts,amsmath,graphicx}

\newcommand{\ket}[1]{|#1\rangle}

\begin{document}

\title{Detecting phonon blockade with photons}
\author{Nicolas Didier}
\affiliation{NEST, Scuola Normale Superiore and Istituto di Nanoscienze - CNR, Pisa, Italy}
\author{Stefano Pugnetti}
\affiliation{NEST, Scuola Normale Superiore and Istituto di Nanoscienze - CNR, Pisa, Italy}
\author{Yaroslav M.~Blanter}
\affiliation{Kavli Institute of Nanoscience, Delft University of Technology, Lorentzweg 1, 2628 CJ Delft, The Netherlands}
\author{Rosario Fazio}
\affiliation{NEST, Scuola Normale Superiore and Istituto di Nanoscienze - CNR, Pisa, Italy}

\begin{abstract}
Measuring the quantum dynamics of a mechanical system, when few phonons are involved, remains a challenge.
We show that a superconducting microwave resonator linearly coupled to the mechanical mode constitutes a very powerful probe for this scope.
This new coupling can be much stronger than the usual radiation pressure interaction by adjusting a gate voltage.
We focus on the detection of phonon blockade, showing that it can be observed by measuring the statistics of the light in the cavity. 
The underlying reason is the formation of an entangled state between the two resonators.
Our scheme realizes a phonotonic Josephson junction, giving rise to coherent oscillations between phonons and photons as well as a self-trapping regime for a coupling smaller than a critical value.
The transition from the self-trapping to the oscillating regime is also induced dynamically by dissipation.
\end{abstract}
\maketitle

\section{Introduction}

Since the early days of quantum mechanics the crossover to the classical world and the possibility that macroscopic objects could exhibit quantum behavior has attracted continuous interest. 
The tremendous progress in nano-fabrication capabilities have made these questions amenable to experimental testing.
Stimulated by the ideas of Leggett~\cite{Leggett}, macroscopic quantum effects were first explored in Josephson junctions~\cite{MartinisDevoret} and nano-magnets~\cite{awschalom}. 
In the recent past, the field of Nano ElectroMechanical Systems (NEMS) has received much attention as a very promising ground for the investigation of these questions~\cite{SchwabRoukes,Huettel,Teufel,Anetsberger,MartinisCleland} and the observation of a mechanical oscillator in its quantum ground state~\cite{MartinisCleland,Teufel2011} is one of the most important achievements reached so far.

The coupling of a quantum nano-mechanical oscillator to a qubit makes NEMS also suitable systems to explore the physics of circuit/cavity-QED~\cite{cQED}.
Different schemes have been proposed including coupling to Cooper-pair boxes~\cite{Armour02,Martin04,Rabl04,Wei06,Armour08,Hauss08} and phase qubits~\cite{Cleland04,Trees07}. 
Recently, coupling to a Cooper-pair box has been realized experimentally~\cite{LaHaye09}. 
Among the numerous interesting aspects of circuit-QED realized with mechanical resonators, here we want to address the phenomenon of {\em phonon blockade}, which was considered recently in Ref.~\onlinecite{Nori}, extending to NEMS the original ideas put forward with photons in cavity-QED systems~\cite{PhotonBlockade}. 
The blockade effect arises because the coherent coupling of the harmonic (photonic/phononic) mode with the (solid-state) atom leads to an effective non-linearity. 
For sufficiently strong coupling, the non-linearity is such that, upon external driving, the number of excitations of the oscillator never exceeds one. 
Observation of phonon blockade in a nano-mechanical oscillator would be a clear evidence of its quantum nature. 

\begin{figure}[b]
\centering
\includegraphics[width=\columnwidth]{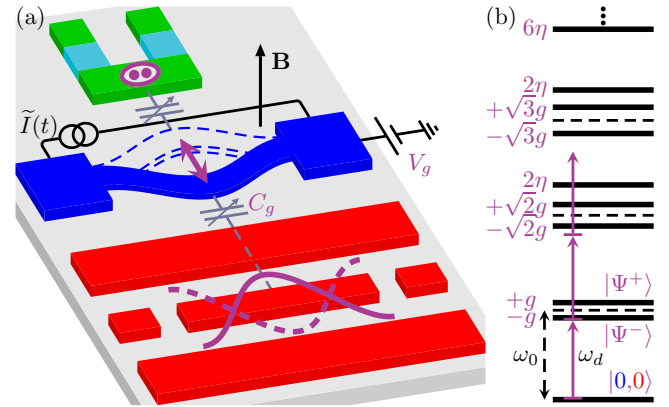}
\caption{(a)~Scheme of the system. 
A mechanical resonator is capacitively coupled to an artificial atom to induce a non-linearity and to a superconducting microwave resonator to detect phonon blockade.
(b)~Schematics of the energy spectrum of the system for $\eta\gg g$.
The first excited states are the maximally entangled Bell states $\ket{\Psi^\pm}$ and are off resonance with the states with more than one excitation.
When the system is excited at the frequency $\omega_d=\omega_0\pm g$, phonon blockade is observed through the photon statistics.}
\label{figsystemspectrum}
\end{figure}

There is, however, an important challenge which needs to be tackled towards the demonstration of phonon blockade: its detection. 
The motion of a mechanical oscillator close to its ground state is tiny and strong amplification of the transduced signals must be applied. 
On the other hand, amplification will inevitably add classical correlations to the signal, thus disguising quantum correlations which are usually needed for demonstrating phonon blockade.

In this article we show that these problems may be overcome if a Superconducting Microwave Resonator (SMR) is coupled linearly to the mechanical oscillator for transducing its motion into an electric signal. SMRs have proven to be nearly ideal quantum oscillators with easily tunable quality factors and they can be efficiently coupled to coherent quantum circuits~\cite{WallraffSchuster,Bishop,Blais,Wallquist,peano}. Our analysis is amenable to experimental verification. Recently, good coupling of a nano-mechanical resonator and a SMR has been demonstrated~\cite{Regal,Schwab,Hertzberg,Teufel2011}. Moreover, despite the difficulties in dealing with a microwave field at the level of a single photon, recently it has been shown that time correlation functions for the cavity field can be measured accurately~\cite{Wallraff,Schoelkopf}.
Very recently, microwave photon blockade has been realized and measured experimentally~\cite{Wallraffblockade}. 
Our detection scheme takes fully advantage of the experimental capabilities of the SMR to probe the quantum state of the mechanical resonator with a linear coupling.

The hybrid phonon-photon system that we analyze in this work goes beyond a mere detection scheme for the phonon blockade. 
In the second part we present the properties of this {\em phonotonic junction}.
For a mechanical oscillator initially driven in an exited state we show that two regimes exist, depending on the ratio between the anharmonicity and the coupling compared to a critical value.
For a small coupling, the phonons are trapped in the mechanical oscillator whereas coherent oscillations between the phonons and the photons appear when increasing the coupling, leading to quantum revivals in the photon statistics.
We show that this transition can be induced by the dissipation and take place during the temporal evolution of the system.

The paper is organized as follows. In the next Section we define the model for the system depicted in Fig.\ref{figsystemspectrum} and introduce the 
equations governing its dynamics in the presence of damping. In Section~\ref{phonoblockade} we discuss how to detect phonon blockade by means of 
a measurement of the photon statistics. In Section~\ref{phonotonic} we present the properties of this phonotonic junction with the transition between a self-trapping regime and a regime with coherent oscillations.
We summarize our results in the concluding Section.

\section{The Model}

The system we have in mind is depicted in Fig.~\ref{figsystemspectrum}. 
It is a mechanical resonator coupled capacitively to an artificial atom and a SMR. 
The non-linearity, leading to phonon blockade, is induced by coupling the oscillator to a superconducting qubit. 
In the case of a Cooper-pair box, one gets the Jaynes-Cummings model. 
A Cooper-pair box molecule~\cite{Milburn} may be considered as well in order to increase the non-linearity. 
In the present work, however, we do not deal with a specific choice of the superconducting nano-circuit; the important ingredient is the generation of the Kerr Hamiltonian proportional to $a^\dag a^\dag a a$~\cite{Jacobs}, where $a$ and $a^\dag$ are the phonon annihilation and creation operators of the mechanical oscillators. 
After the adiabatic elimination of the qubit, the Non-linear Mechanical Resonator (NMR) is described by the following effective Hamiltonian
\begin{equation}
H_\mathrm{NMR}=\hbar\omega_ra^\dag a + \hbar\eta a^\dag a^\dag a a.
\end{equation}
The mechanical resonator is supposed to be in the quantum regime (the bare frequency is in the GHz range).
For a strong coupling of hundreds of MHz, the Kerr non-linearity strength $\eta$ is of the order of MHz. The coupling between the NMR and the SMR is obtained from circuit theory. 
The cavity is modeled by an array of $LC$ circuits~\cite{Wallquist}. 
The Hamiltonian of the fundamental mode of the SMR is $H_\textrm{SMR}=\hbar\omega_c b^\dag b$, where $\omega_c/2\pi$ is the mode frequency and $b$ ($b^\dag$) the corresponding photon annihilation (creation) operator.

The mechanical resonator is kept at a fixed potential $V_g$ with respect to ground and the coupling is realized through a localized capacitance $C_g=C_g^0+(a+a^\dag)C_g^1$, resulting in two coupling terms in the Hamiltonian: a radiation pressure term $C_g^1V_\mathrm{r.m.s.}^2(a+a^\dag)b^\dag b$ and a linear term $C_g^1V_g V_\mathrm{r.m.s.}(a+a^\dag)(b+b^\dag)$, coming from the electrostatic energy of the coupling capacitor ($V_\mathrm{r.m.s.}$ is the root mean square of the zero-point voltage fluctuation of the SMR at the position of the coupling capacitance). 
These two terms can have very different orders of magnitude for typical values of $V_g$ and $V_\mathrm{r.m.s.}$~\cite{Tian09}.
We focus on the case where the gate voltage is much larger than the root mean square of the voltage quantum fluctuations inside the SMR, and thus neglect radiation pressure. 
In the rotating wave approximation, the coupling reads
\begin{equation}
H_\mathrm{int}=\hbar g \left( a^\dag b + a b^\dag \right),
\end{equation}
where $g=C_g^1V_g V_\mathrm{r.m.s.}/\hbar$.
For realistic parameters, the coupling may be of the order of MHz.
The mechanical resonator is driven by a weak oscillating current, at the frequency $\omega_d/2\pi$, in the presence of a static magnetic field perpendicular to the plane of the circuit. 
This driving is modeled by the Hamiltonian 
\begin{equation}
H_\mathrm{drv}=\hbar\epsilon\left(a^\dag e^{-i\omega_dt}+a e^{i\omega_dt}\right),
\end{equation}
where the strength $\epsilon$ is proportional to the current amplitude and the magnetic field.
Driving the mechanical oscillator can in principle act as a direct driving also for the SMR through the capacitive and inductive couplings.
However, the resulting driving is negligible.
Indeed, on the one hand because the displacement is tiny the electromotive force induced in the SMR is well below $V_\mathrm{r.m.s.}$ and on the other hand for spatially separated resonators the mutual inductance is suppressed.
The total Hamiltonian of the system reads $H=H_\mathrm{NMR}+H_\mathrm{SMR}+H_\mathrm{int}+H_\textrm{drv}$.
We choose the working point at $\omega_r=\omega_c\equiv\omega_0$.
The mode frequency of the cavity can be tuned by adding a SQUID at the end of one arm to change the boundary condition~\cite{wilson}.
In the rotating frame of the driving, the total Hamiltonian reads
\begin{multline}
H=\hbar(\omega_0-\omega_d)(a^\dag a+b^\dag b) + \hbar\eta a^\dag a^\dag aa\\+ \hbar g(a^\dag b+ab^\dag) + \hbar\epsilon(a^\dag+a).
\end{multline}
A linear coupling between driven non-linear oscillators can also be obtained in cavity-QED~\cite{leonski}.

The finite lifetime of the phonons and the photons is taken into account through the Lindblad operators $L_r$ and $L_c$ of the resonator and the cavity, respectively~\cite{ScullyCarmichael} ($\rho$ is the density matrix of the whole system)
\begin{subequations}
\begin{align}
L_r\rho&=\tfrac{1}{2}\gamma_r\left(2a\rho a^\dag-a^\dag a\rho-\rho a^\dag a\right),\\
L_c\rho&=\tfrac{1}{2}\gamma_c\left(2b\rho b^\dag-b^\dag b\rho-\rho b^\dag b\right),
\end{align}
\end{subequations}
where the damping rates $\gamma_{r,c}=\omega_{r,c}/Q_{r,c}$ are the inverse of the phonon and photon lifetime and are defined by the quality factors $Q_{r,c}$.
The dynamics of the system is then governed by the master equation
\begin{equation}
\partial_t\rho(t)=\frac{1}{i\hbar}\left[H,\rho(t)\right] + L\rho(t),
\label{meq}
\end{equation}
where $L=L_r+L_c$ is the total Lindbladian.

In principle, some noise is introduced by the voltage source used to keep the mechanical resonator at $V_g$. 
However, following Ref.~\onlinecite{Makhlin01}, 
the dominant Lindblad operator corresponding to this source of noise for the case of a Markovian environment is found to be of the order of $\eta g R C_g^1 V_\mathrm{r.m.s.}/V_g$ ($R$ is the internal resistance of the voltage source), which is negligible due to the very small value of $gRC_g^1$.
In the case of slow voltage fluctuations, one can also neglect noise effects, as we discuss below.

\section{Detection of phonon blockade}
\label{phonoblockade}

\begin{figure}[t]
\centering
\includegraphics[height=6cm]{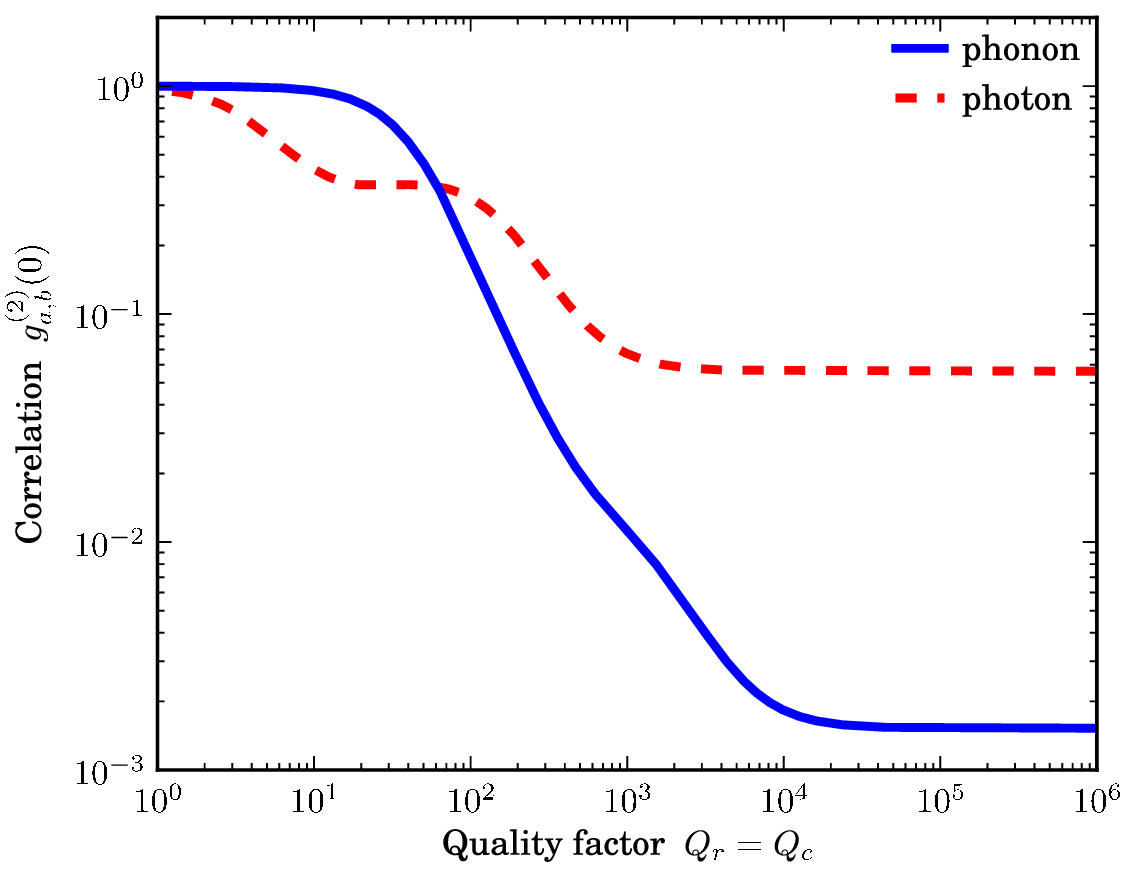}
\caption{(Color online) 
Second order correlation functions against the quality factors.
The phonotonic Josephson junction works for quality factors down to $Q_{r,c}\sim10^3$, where the damping rates are comparable to the coupling strength.}
\label{figQ}
\end{figure}

Blockade is possible only if the non-linearity of the energy spectrum is larger than the state linewidth, 
namely $\eta,g>\gamma_{r,c}$. 
This condition imposes the quality factors to be at least several thousand, which is within the experimental capabilities. 
Throughout the present work we chose the following parameters for the numerical simulations: $\omega_0/2\pi=1\,\mathrm{GHz}$, $\eta/2\pi=10\,\mathrm{MHz}$, $g/2\pi=1\,\mathrm{MHz}$, $Q_r=10^5$, and $Q_c=10^6$. 
Fig.~\ref{figQ} shows that the results are still valid for quality factors $Q_{r,c}=3000$, when the damping becomes of the order of the coupling.
Moreover, the NMR is driven at the resonance $\omega_d=\omega_0-g$ with the amplitude $\epsilon/2\pi=0.1\,\mathrm{MHz}$.
By analyzing the time traces of the photon and phonon populations (not shown here) it is possible to note that the cavity closely resembles the dynamics of the NMR, with equal steady state phonon and photon numbers close to $0.25$.
The population of the state with two phonons or two photons is strongly suppressed, implying both phonon and photon blockade. 
In the rest of the paper, we will show by solving Eq.~\eqref{meq}, that by means of the detection of photon correlations it is possible to extract unique information on the phonon statistics.

\begin{figure}[t]
\centering
\includegraphics[height=6cm]{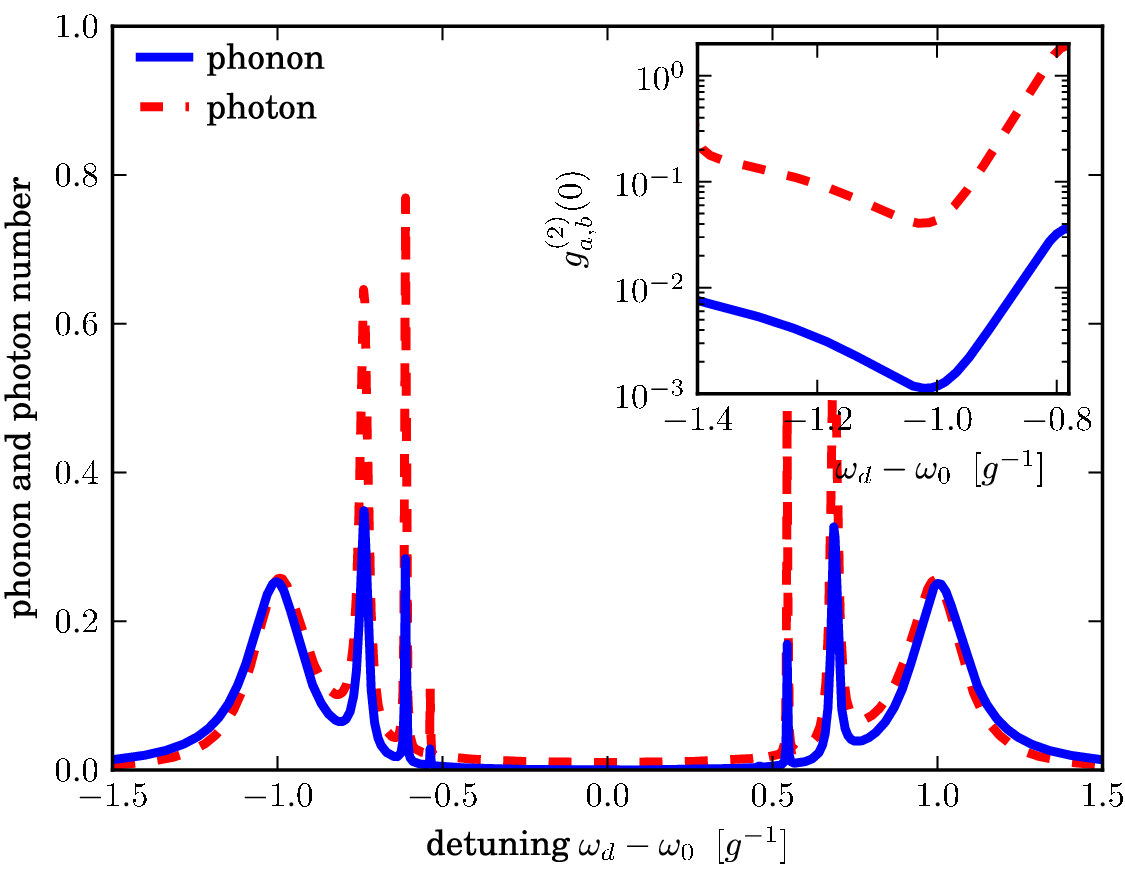}
\caption{(Color online) 
Spectroscopy of the energy spectrum.
Steady state phonon number (solid line) and photon number (dashed line) as a function of the driving frequency.
The different peaks correspond to the excitation of the states with one phonon $\ket{1,n_c}$.
Inset: Second order correlation functions at coinciding times for the phonons $g^{(2)}_a(0)$ and for the photons $g^{(2)}_b(0)$.
Anti-bunching occurs at $\omega_d=\omega_0\pm g$ where phonon and photon blockade take place.
Photon bunching occurs when the states with many photons are excited.}
\label{figd}
\end{figure}

In order to ascertain the accuracy of the proposed detection scheme, it is not sufficient to see a correlation in the populations. 
We analyze the statistics of the excitations by means of the second order correlation function~\cite{MandelWolf}
\begin{equation}
g^{(2)}_y(\tau)=\lim_{t\to\infty}\frac{\langle y^\dag(t) y^\dag(t+\tau) y(t+\tau) y(t) \rangle}{\langle y^\dag(t) y(t)\rangle^2},
\end{equation}
where $y=a$ or $b$.
The value of $g^{(2)}_{a,b}(\tau)$ is comprised between $0$ and $2$ and tends towards unity for long time difference, where the coherence is lost.
The value at coinciding times $\tau=0$ reflects the statistics of the field: 
a value of $g^{(2)}_a(0)\ll1$ corresponds to antibunching, and serves as the signature of phonon blockade.

\begin{figure}[t]
\centering
\includegraphics[height=6cm]{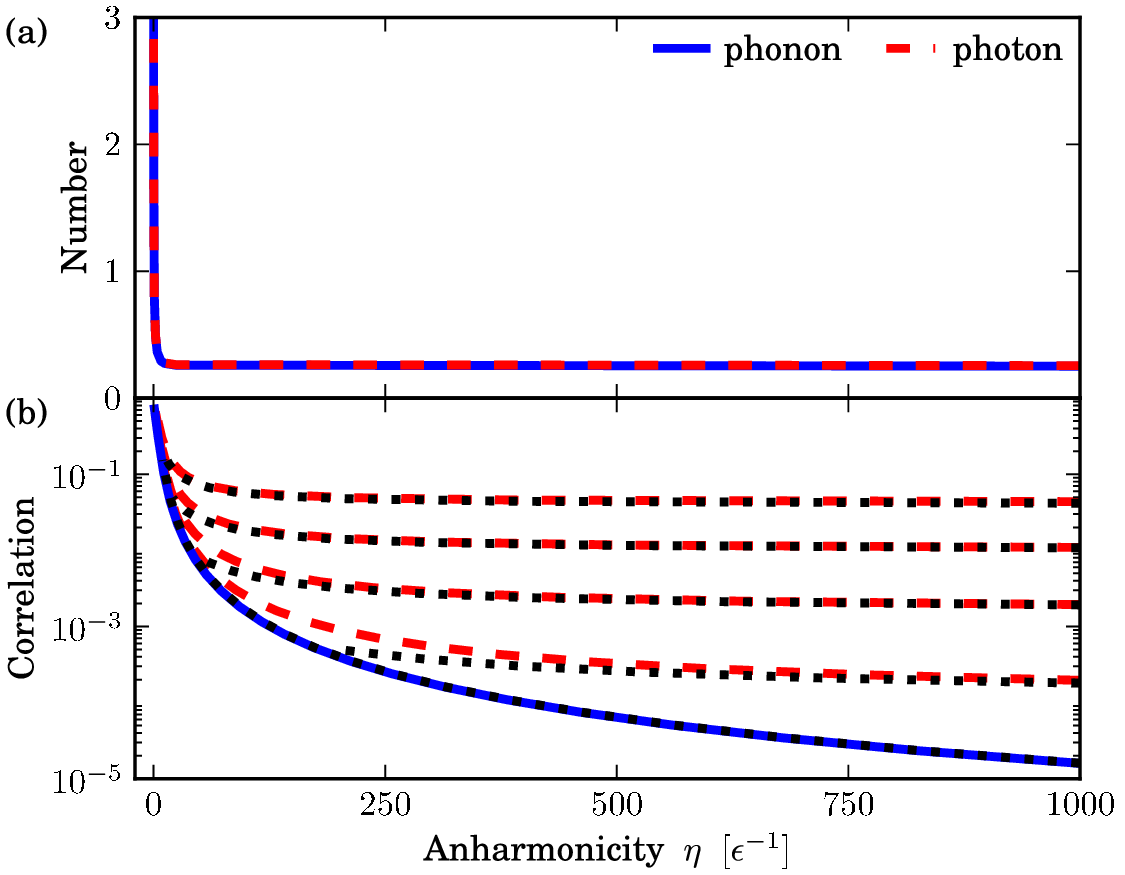}
\caption{(Color online) 
Steady state phonon and photon numbers (a) and second order correlation functions (b) as a function of the Kerr non-linearity.
The Bell state $\ket{\Psi^-}$ ensures a perfect match between the phonon and photon numbers.
Many excitations are generated when the non-linearity is comparable with the damping rates of the resonators, where the excitations are not anti-bunched anymore.
The correlators are plotted for different values of the coupling,
from top to bottom, $g/2\pi=1,\ 2,\ 5,\ 20\,\mathrm{MHz}$ respectively.
The correlation of the NMR is essentially coupling-independent while for large $\eta$ the correlation of the SMR saturates to a constant value $(2\epsilon/g)^2$.
The corresponding solutions Eq.~\eqref{g2largeeta} in the limit $\eta\gg g$ are plotted in dotted lines.}
\label{figK}
\end{figure}

To understand how to induce and detect phonon blockade, we look at the energy spectrum of the undriven Hamiltonian $H$ in the Fock basis $\ket{n_r,n_c}$, where $n_{r,c}$ is the phonon and photon number respectively.
The total number of excitation $a^\dag a+b^\dag b$ being conserved, the spectrum can be decomposed on the subspaces defined by a given number $n=n_r+n_c$ of excitations $\{\ket{k,n-k},\ k=0,\dots,n\}$. 
For one excitation $n=1$, the eigenstates are the maximally entangled Bell states
\begin{equation}
\ket{\Psi^\pm}=\tfrac{1}{\sqrt{2}}\left(\ket{0,1}\pm\ket{1,0}\right),
\end{equation}
with the energy $\hbar\omega_0 \pm \hbar g$.
If there is a small detuning between the two resonators, the eigenstates are rotated by an angle $(\omega_r-\omega_c)/4g$.
For higher excitation numbers the ladder structure depends on the ratio $\eta/g$ between the anharmonicity and the coupling.
In the limit of a strong non-linearity $\eta\gg g$, the spectrum is composed of two entangled states $\ket{0,n}\pm\ket{1,n-1}$ at $\omega_0\pm\sqrt{n}g$ and $n-1$ pure states $\ket{2,n-2},\dots,\ket{n,0}$ located at $\hbar\omega_0+m(m-1)\eta$, see Fig.~\ref{figsystemspectrum}~(b).
This non-linear spectrum allows for excitation blockade, since the energy of the state $\ket{\Psi^\pm}$ is not resonant with higher states.
If the system is excited at the frequency $\omega_d=\omega_0\pm g$, only one excitation is created, symmetrically shared between the NMR and the SMR.
The eigenstate being the maximally entangled Bell state $\ket{\Psi^\pm}$, the photons have the same dynamics as the phonons and the cavity constitutes consequently a reliable measurement device to detect the state of the resonator through the photon statistics (see Fig.~\ref{figK}).

The energy spectrum can be probed with the response of the system to the driving current when the driving frequency is tuned, as depicted in Fig.~\ref{figd}. 
In order to excite states with one phonon such as $\ket{1,n-1}$, the driving frequency is fixed to $\omega_d=\omega_0\pm g/\sqrt{n}$. 
These values correspond to the peaks in the excitation numbers of Fig.~\ref{figd}. 
Compared with the dependence of the $g_{a,b}^{(2)}(0)$ on $\omega_d$ (see the inset), it shows that blockade occurs at $\omega_d-\omega_0=\pm g$ where the second order correlation function is minimized. 
This minimum protects the blockade phenomenon against slow fluctuations of the gate voltage $V_g$, or equivalently the coupling $g$.
Indeed, if the driving is kept at a fixed frequency $\omega_d=\omega_0\pm g$ and one looks at the dependence of $g_{a,b}^{(2)}(0)$ on a coupling $\tilde{g}(t)$ fluctuating around the value $g$, one obtains a behavior analogous to the inset of Fig.~\ref{figd}, namely a minimum at $\tilde{g}=g$.
The averaged value of the second order correlation function over a Gaussian distribution of coupling strengths is not affected and both phonon and photon blockade is thus insensitive to slow gate voltage fluctuations at first order.
For higher values of $n$, the photon number increases and the second order correlation function of the cavity $g_b^{(2)}(0)$ tends to 2, indicating photon bunching.

When the anharmonicity is very large, the NMR behaves like a two-level system and can be described 
by replacing the operators $a$ and $a^\dag$ by the the ladder operators $\sigma_-=\ket{0}\langle1|$ and $\sigma_+=\ket{1}\langle0|$, respectively. 
The Hamiltonian is then reduced to an effective Jaynes-Cummings model, $H_\mathrm{JC}=\hbar g(\sigma_-b^\dag+\sigma_+b)$.
In lowest order in $g/\eta$, the asymptotic expression of the second order correlation functions is given by
\begin{equation}
g^{(2)}_a(0)=(4\epsilon/\eta)^2,\quad g^{(2)}_b(0)=(2\epsilon/g)^2(1+4g/\eta).
\label{g2largeeta}
\end{equation}
The comparison with the numerical results is presented in Fig.~\ref{figK}, where the correlation functions are plotted as a function of the anharmonicity for different values of the coupling.
For a sufficiently large anharmonicity the properties of the SMR become $\eta$-independent with a strong reduction of the phonon correlation function $g_a^{(2)}(0)$. 
Photon blockade is enhanced when the coupling increases.
In the opposite limit of a small anharmonicity, the driving generates many excitations.
The transition from photon anti-bunching to photon bunching is also observed when, for a fixed anharmonicity, the coupling decreases or the driving amplitude increases. 
The former is due to the degeneracy of the Bell states for small coupling $g\lesssim\gamma_{r,c}$ and the latter is because of the effective level broadening due to the driving.

\section{The phonotonic junction}
\label{phonotonic}

\begin{figure}[t]
\centering
\includegraphics[height=6cm]{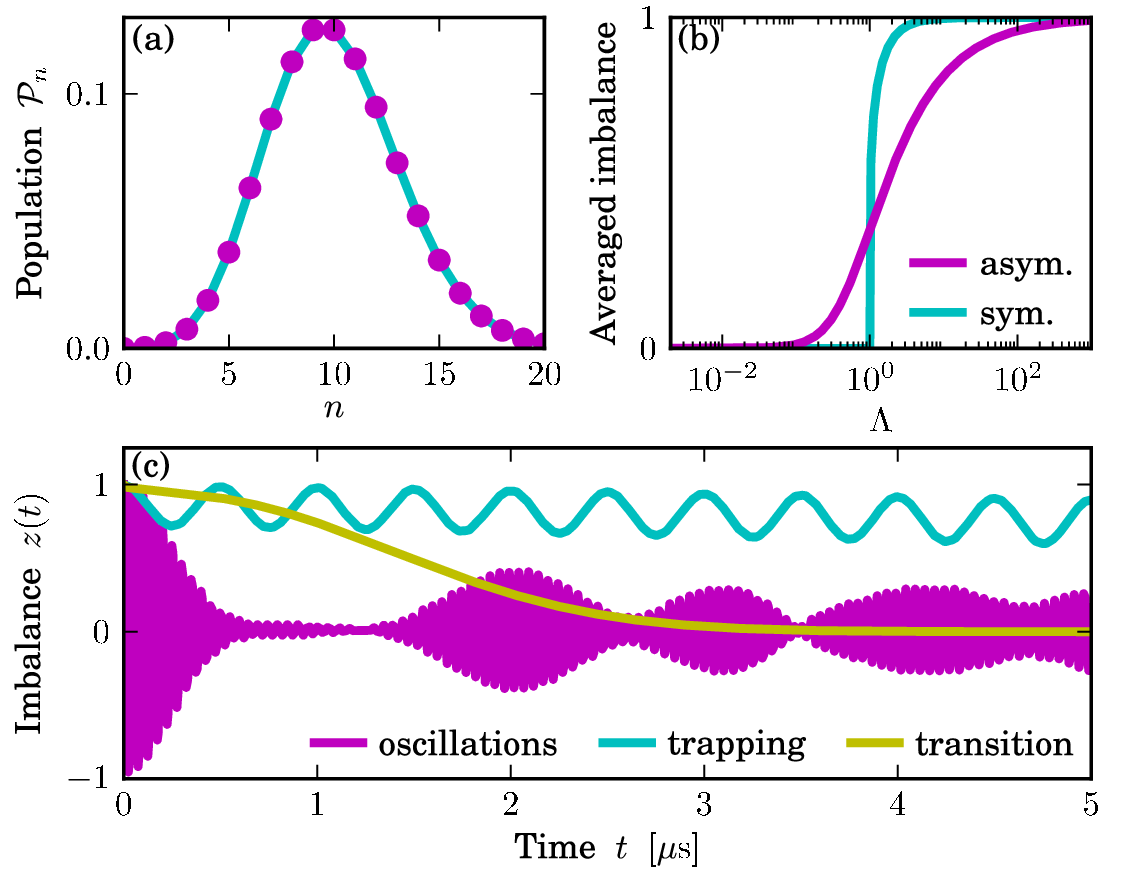}
\caption{(Color online) 
(a) Population distribution of the mechanical oscillator after being driven to a state with 10 phonons (points).
The statistics is compared to a coherent state with the same mean value (line).
(b) Average value of the imbalance $z$ as a function of the parameter $\Lambda$ from the classical dynamics without dissipation, starting from the initial state $z(0)=1$.
The symmetric junction corresponds to interacting photons with a strength $\eta$.
(c) Time evolution of the population imbalance in the self-trapping regime ($\eta/2\pi=10\,\mathrm{MHz}$ and $g/2\pi=1\,\mathrm{MHz}$), the oscillating regime ($\eta/2\pi=1\,\mathrm{MHz}$ and $g/2\pi=10\,\mathrm{MHz}$).
The dynamical transition is observed with a symmetric junction, where the photons also interact with a strength $\eta/2\pi=10 \,\mathrm{MHz}$. The coupling is equal to $g/2\pi=10 \,\mathrm{MHz}$ and the quality factors are $Q_{r,c}=10^4$.
}
\label{figjunction}
\end{figure}

For an isolated mechanical oscillator ($\eta=g=\gamma_r=0$) initially in its ground state $\ket{0}_r$, the resonant driving ($\omega_d=\omega_r$) generates a coherent state $\ket{\alpha=-i\epsilon t}_r=\exp[-i\epsilon t(a^\dag+a)]\ket{0}_r$ after a time $t$ (see Fig.~\ref{figjunction}~(a)).
In the presence of dissipation, at zero temperature, the steady state is the coherent state $\ket{\alpha=-i2\epsilon/\gamma_r}_r$~\cite{ScullyCarmichael}.
The synthesized coherent state $\ket{\alpha}_r$, with a tunable phonon number $|\alpha|^2$, will be used as the initial state in the following.
Once the phononic state is prepared, the driving is turned off while the coupling to the qubit and the SMR are switched on.
The latter can be performed by tuning the qubit frequency closer to the resonance and putting the gate voltage $V_g$ on, respectively.
The system is then governed by a two-sites Bose-Hubbard like Hamiltonian $H=\hbar\omega_0(a^\dag a+b^\dag b) + \hbar\eta a^\dag a^\dag aa + \hbar g(a^\dag b+ab^\dag)$.
This is similar to the Bosonic Josephson junction, realized with a cloud of cold atoms in a double well potential~\cite{Albiez}.
Since only one of the two resonators is non-linear, we are dealing with an asymmetric junction.
The symmetric case, where the photons are also interacting, can be obtained by adding a qubit in the SMR to generate a non-linearity~\cite{schuster}.
To present the properties of this junction, we start with a classical description in the absence of dissipation.

The dynamics of the phonotonic junction can be described in terms of the imbalance $z=\langle n_r-n_c\rangle/\langle n_r+n_c\rangle$ 
between the phonons and the photons and the phase difference $\varphi=\arg\langle a^\dag b\rangle$ between the two resonators.
In the absence of dissipation, the total number of particles $\langle n_r+n_c\rangle$ is conserved, equal to $|\alpha|^2$.
The classical equations of motion for $\langle n_r\rangle$, $\langle n_c\rangle$, and $\langle a^\dag b\rangle$ can be expressed in terms of $z$ and $\varphi$.
The system is then classically governed by the following set of nonlinear differential equations
\begin{align}
\dot{z}(\tau)&=\sqrt{1-z^2(\tau)}\sin\varphi(\tau),\label{classicalimbalance}\\
\dot{\varphi}(\tau)&=\Lambda[1+z(\tau)]-\frac{z(\tau)}{\sqrt{1-z^2(\tau)}}\cos\varphi(\tau),\label{classicalphase}
\end{align}
where the time has been rescaled to $\tau=2gt$.
The parameter $\Lambda=\eta\langle n_r+n_c\rangle/2g$ leads to two regimes~\cite{smerzi}.
For large couplings, coherent oscillations take place between the phonons and photons.
When the interaction between phonons is larger than the coupling, the oscillations are frozen and the particles are self-trapped.
The transition between these two regimes is controlled by the critical parameter $\Lambda_c=2$, corresponding to a critical coupling
\begin{equation}
g_c(t)=\tfrac{1}{4}\eta\langle n_r+n_c\rangle(t).
\end{equation}
These two regimes are presented in Fig.~\ref{figjunction}~(b).
The time average of the imbalance vanishes in the oscillating regime $\Lambda<\Lambda_c$ and tends to unity in the self-trapping regime $\Lambda>\Lambda_c$.
The imbalance is also presented for a symmetric junction, where the photons are also interacting with a strength $\eta$.
For a symmetric junction, the imbalance follows the dynamics of Eq.~\eqref{classicalimbalance} while concerning the phase the first term in the right-hand side of Eq.~\eqref{classicalphase} has to be replaced by $2\Lambda z(\tau)$.
This gives rise to a sharp transition at $\Lambda=1$.

In the presence of dissipation, the critical coupling decreases with the characteristic rate $\gamma_{r,c}$.
If the coupling is initially larger than the critical coupling $g_c(0)$, the system starts in the self-trapping regime.
After a time $t_0\sim\ln[g_c(0)/g]/\gamma_{r,c}$, the coupling becomes larger than the critical coupling $g_c(t_0)$ and coherent oscillations take place between phonons and photons.
Quantum revivals can then be seen in the cavity.
This dynamical transition induced by dissipation is sharper when the photons are also interacting\cite{gerace}.
The transition is presented in Fig.~\ref{figjunction}~(c) for a symmetric junction.

\section{Conclusion}

In conclusion, we have shown that coupling a SMR to a NMR is a powerful tool to detect phonon blockade and generate entanglement between phonons and photons.
The main reason for the accurate detection when few phonons are involved is the formation of Bell states between the two resonators, ensuring a perfect match between the phonon dynamics and the photon statistics.
The phonotonic Josephson junction takes advantage of the recent experimental capabilities with microwave photons. 
The simulations, obtained in the framework of the quantum master equations, demonstrate that our proposal is compatible with the current experimental capabilities.
Our new detection scheme realizes a phonotonic Josephson junction and its applications go beyond the blockade regime.
A rich physics stems from this device which can be used for instance to observe phonon lasing through microwave photon lasing or the dynamical Casimir effect if the gate voltage is strongly modulated.
The phonotonic Josephson junction constitutes a building block towards the use of NEMS as quantum buses~\cite{Zoller}.

We acknowledge financial support from EU through the projects QNEMS, SOLID and GEOMDISS.

\end{document}